\documentclass[a4paper,11pt]{article}
\usepackage[usenames]{color}

\usepackage{amsfonts,amssymb}
\usepackage{theorem}
\usepackage{cancel}
\usepackage[dvips]{graphicx}

\def\g5{\gamma_{_\chi}}

\newcommand{\comment}[1]{}

\begin{document}
\comment
{
  \rightline{\Huge gamma\_5\_paper\_v4\_arxiv\_1403.4212v2.tex}
\par
\rightline{IFUM-xxx-FT}
\par
\rightline{Pisa, March 13, 2014}
\par
\rightline{revised Pisa February 27,  2015 at 12.00}
}
\rightline{\sl To Roman Jackiw}
\vskip 1.0 truecm
\Large
\bf
\centerline{Managing $\gamma_5$ in Dimensional Regularization
and ABJ Anomaly
\footnote{\tt This work is supported in part by funds provided by the U.S. Department
of Energy (D.O.E.) under cooperative research agreement \#DE FG02-05ER41360} }
\par
\normalsize
\rm

\rm 
\large
\vskip 1.1 truecm
\centerline{
Ruggero~Ferrari \footnote{e-mail: {\tt ruggferr@mit.edu}}}
\normalsize
\medskip
\begin{center}
Center for Theoretical Physics\\
Laboratory for Nuclear Science\\
and Department of Physics\\
Massachusetts Institute of Technology\\
Cambridge, Massachusetts 02139\\
and\\ INFN, Sezione di Milano\\
via Celoria 16, I-20133 Milano, Italy\\
(MIT-CTP4517 March 2014)
\end{center}
%
%

\normalsize
\bf
\centerline{Abstract}

\rm
\begin{quotation}
An integral representation  is proposed for
the trace involving $\gamma_5$ in
dimensional regularization. Lorentz covariance is preserved.
ABJ anomaly naturally follows. The Local
Functional Equation associated to x-dependent chiral transformations
is verified.

\end{quotation}
\newpage
\section{Introduction}
\label{sec:intr}
%
In dimensional regularization (Ref. \cite{'tHooft:1972fi},\cite{Bollini:1972ui}
and \cite{Cicuta:1972jf})
$\gamma_5$ has always been a very difficult object
to deal with. Many important contributions to the topics are
present in the literature. We provide an uncommented list 
of works  
{
\cite{Rosenberg:1962pp}, \cite{'tHooft:1972fi},
\cite{Breitenlohner:1977hr}-\cite{Tsai:2009hp},
}
which is far from being complete.
\par
In this paper an integral representation
of the trace involving $\gamma_5$ is suggested
\begin{eqnarray}&&
Tr(\not\! p_1\dots \not\! p_{_N}\gamma_5)
\nonumber\\&&  
=i^{\frac{D(D-1)}{2}}K \int d^{^D}\chi\, d^{^N} \bar c\, 
\exp \Big(\sum_{\mu=1}^D\sum_{i=1}^N\bar c_i p_{i\mu}\chi_\mu 
+\sum_{i<j}^N\bar c_i( p_{i} p_j) \bar c_j\Big ),
\label{int.repr.0.0}
\end{eqnarray}
where $\bar c_i ~~ \chi_\mu$ are Grassmannian 
variables. $K$ is a real parameter related to the normalization of
the trace.
\par
The strategy is the following: {for generic \underline{integer}
value of $D$ we investigate on the properties
of the integral representation proposed in (\ref{int.repr.0.0}).
In particular a) we show that the
Clifford's algebra  of the gamma's is obeyed; 
b) the cyclicity of the trace is proved if no $\gamma_5$ is present
(more precisely: no $\chi$ integration);
c) if the  $\gamma_5$ is present  we require  cyclicity of the trace 
operation. 
}
\par
Once these results are established, 
{we assume that the amplitudes can be continued in $D$
to become a meromorphic function in the complex plane.
Then} we can 
dimensionally regularize  any
Feynman amplitude and use those manipulations that
are consistent with the regularization in eq. (\ref{int.repr.0.0}).
\par\noindent
{In the present paper we also} provide a closed formula for the evaluation of the trace.
\par
We test our proposal with the evaluation of the axial anomaly.
ABJ result 
\cite{Adler:1969gk} and \cite{Bell:1969ts} is obtained in a very natural and elegant
way. 
\par
Moreover a Local Functional Equation (LFE)
\cite{Ferrari:2005ii}-\cite{Bettinelli:2007kc} is derived and checked
at one loop level. This fact is noteworthy: the LFE
doesn't suffer of anomalies.
\par
In order to make easy the notations we use Euclidean metric 
for indices and we change the name ($\chi$ for {\sl chiral})
\begin{eqnarray}
\gamma_5  \to \gamma_\chi\, .
\label{intr.1}
\end{eqnarray}
%
%
\section{Integral Representation of Trace of Gamma's Product
}
\label{sec:int.rep}
We consider the trace of a generic product of gamma's 
\cite{E.R.Caianiello:1952ww}-\cite{Veltman:1988au},
where the indices are saturated by vectors $p_{j\mu}$
($\not\! p_j$ is a standard notation for $\gamma_\mu p_{j\mu}$).
The discrete index $j$ runs over the set of
integers $\{1,\dots, N\}$,{ while the ``component index'' $\mu$
is an element of the set $\{1,\dots, D\}$ (of $\mathbb N$).} Our aim is to
find an integral representation for the trace with and 
without $\gamma_\chi$. 
\par
We prove the integral representation by
showing the  validity of the gamma's algebra (Clifford)
(with no mention to the dimension $D$). For integer $D$
we prove the property of cyclicity and evaluate the algebra of
$\gamma_\chi$. 
\par
We use the standard properties of integration on Grassmannian real
coordinates
\begin{eqnarray}&&
\int d \bar c =0
\nonumber\\&&
\int d \bar c \,\,\bar c = 1
\nonumber\\&&
\bar c' \equiv -\bar c,  \quad \Longrightarrow \int d \bar c'\,\, \bar c' = 1.
\label{int.repr.1.-1}
\end{eqnarray}
%

\subsection{If $N$ is even and no $\gamma_\chi$ is present} 
The trace can be written
in terms of an integral over a set of Grassmannian variables
$\bar c_j$
\begin{eqnarray}&&
Tr(\not\! p_1\dots \not\! p_{_N})
=K \int d^{^N} \bar c\, 
\exp \Big(
\sum_{i<j}^N\bar c_i( p_{i} p_j) \bar c_j\Big ) 
\nonumber\\&&
=K \int d^{^N} \bar c\, 
\exp \Big(\frac{1}{2}
\sum\bar c_i h_{ij} \bar c_j\Big ) 
,
\label{int.repr.1.0}
\end{eqnarray}
where
\begin{eqnarray} 
 h_{ij}=-h_{ji}\equiv
( p_{i} p_j) \quad {\rm for}~ i<j 
.
\label{int.repr.1.0.0}
\end{eqnarray}
\par
In the expansion of the exponential only the monomials
containing all $\bar c$ (and only once) yield non-zero
result under integration. The monomials have the form
\begin{eqnarray}
[\bar c_{i_1}( p_{i_1} p_{j_1}) \bar c_{j_1}]\,\,
[\bar c_{i_2}( p_{i_2} p_{j_2}) \bar c_{j_2}]\,\, \dots
\, 
\label{int.repr.1.0.1}
\end{eqnarray}
where $\{i_1,j_1,i_2,j_2\dots \}$ is any permutation $\mathbb P$ of $\{1,2,\dots,N\}$
conditioned by
\begin{eqnarray}&&
i_1<i_2<i_3 \dots
\nonumber\\&&
i_1<j_1,\,\, i_2<j_2, \,\, i_3<j_3,\,\,  \dots\,\,\, .
\, 
\label{int.repr.1.0.2}
\end{eqnarray}
The integration over $d\bar c_{_N}\dots d\bar c_1$ yields
\begin{eqnarray}
\delta_{\mathbb P}
[( p_{i_1} p_{j_1})]\,\,
[( p_{i_2} p_{j_2})]\,\,[( p_{i_3} p_{j_3})]\,\, \dots
\, ,
\label{int.repr.1.0.3}
\end{eqnarray}
where $\delta_{\mathbb P}$ is the signature of the permutation. 
\par
We prove the validity of eq. (\ref{int.repr.1.0}) by showing
that the algebra of the gamma's and cyclicity are obeyed.
\subsection{The Algebra for  even $N$ (and no $\gamma_\chi$)}
\label{sec:even_al}
We evaluate the algebra of the gamma's. 
\par
We start with 
\begin{eqnarray}&&
Tr(\{\not\! p_1,\not\!p_2\}\not\!p_3\dots \not\! p_{_N})
=  K\int d^{^N} \bar c\, 
\exp \Big(
 \sum_{i<j}^N\bar c_i( p_{i} p_j) \bar c_j\Big ) 
\nonumber\\&&
+ K\int d^{^N} \bar c\, 
\exp \Big(\bar c_1 (p_2p_1) \bar c_2  +\sum_{j=3}^N\bar c_1 (p_2p_j) \bar c_j 
+\sum_{j=3}^N\bar c_2(p_1p_j)\bar c_j
\nonumber\\&&
+\sum_{i=3,i<j}^N\bar c_i( p_{i} p_j) \bar c_j\Big ).
\label{int.repr.1.1}
\end{eqnarray}
In the second integral we rename $\bar c_1 \leftrightarrow \bar c_2$.
The measure changes sign and therefore
\begin{eqnarray}&&
Tr(\{\not\! p_1,\not\!p_2\}\not\!p_3\dots \not\! p_{_N})
= K\int d^{^N} \bar c\, 
\exp \Big(
\sum_{i<j}^N\bar c_i( p_{i} p_j) \bar c_j\Big ) 
\nonumber\\&&
- K\int d^{^N} \bar c\, 
\exp \Big(-\bar c_1 (p_2p_1) \bar c_2  +\sum_{j=3}^N\bar c_2 (p_2p_j) \bar c_j 
+\sum_{j=3}^N\bar c_1(p_1p_j)\bar c_j
\nonumber\\&&
+\sum_{i<j=3}^N\bar c_i( p_{i} p_j) \bar c_j\Big ) 
\nonumber\\&&
=K\int d^{^N} \bar c\, 
\exp \Big(
\sum_{i<j}^N\bar c_i( p_{i} p_j) \bar c_j\Big )
\Bigg(1-\exp \Big(-2\bar c_1 (p_1p_2) \bar c_2\Big) \Bigg) 
\nonumber\\&&
=K\int d^{^N} \bar c\,\Big[2
\bar c_1 (p_1p_2) \bar c_2  \Big]
\exp \Big(
\sum_{2<i<j}^N\bar c_i( p_{i} p_j) \bar c_j\Big )
.
\label{int.repr.1.1.1}
\end{eqnarray}
By integrating over  $d\bar c_2 d\bar c_1$ we finally
get
\begin{eqnarray}
Tr(\{\not\! p_1,\not\!p_2\}\not\!p_3\dots \not\! p_{_N})
= 2 (p_1p_2) Tr(\not\! p_3\dots \not\! p_{_N}) .
\label{int.repr.1.2}
\end{eqnarray}
Thus the algebra (Clifford)  of the gamma's is in agreement with the 
representation in eq. (\ref{int.repr.1.0})
\begin{eqnarray}
\{\not\!p,\not\!q\} = 2 (pq) {\mathbb I }
\label{int.repr.1.2.0}
\end{eqnarray}
with
\begin{eqnarray}
Tr\big( {\mathbb I }\big) = K.
\label{int.repr.1.2.1}
\end{eqnarray}
\subsection{Cyclicity for  even $N$ (and no $\gamma_\chi$)}
\label{sec:even_cycl}
We check now cyclicity
\begin{eqnarray}&&
Tr(\not\! p_2\dots \not\! p_{_N}\not\! p_1)
=K \int d \bar c_{_N} \dots d\bar c_1\, 
\exp \Big(
\sum_{i=1,i<j}^{N-1}\bar c_i( p_{i+1} p_{j+1}) \bar c_j
\nonumber\\&&
+\sum_{i=1}^{N-1}\bar c_i( p_{i+1} p_{1}) \bar c_N \Big ) .
\label{int.repr.1.3}
\end{eqnarray}
We rename 
\begin{eqnarray}
\bar c_j \to \bar c_{j+1}\quad \bar c_{_N} \to  \bar c_1
\label{int.repr.1.3.0}
\end{eqnarray}
and get
\begin{eqnarray}&&
Tr(\not\! p_2\dots \not\! p_{_N}\not\! p_1)
=K \int d \bar c_1 d \bar c_{_N} \dots  d\bar c_2\, 
\exp \Big(
\sum_{i=2,i<j}^N\bar c_i( p_{i} p_{j}) \bar c_j
\nonumber\\&&
-\sum_{j=2}^N\bar c_1( p_{1} p_{j}) \bar c_j \Big ).
\label{int.repr.1.3.1}
\end{eqnarray}
Now the minus sign emerging by the ordering of the measure is compensated by 
the change of sign of the $p_1$- dependent terms in the exponential. 
Finally we obtain the identity
\begin{eqnarray}
Tr(\not\! p_2\dots \not\! p_{_N}\not\! p_1)
=
Tr(\not\! p_1\not\! p_2\dots \not\! p_{_N}) .
\label{int.repr.1.4}
\end{eqnarray}
Both eqs. (\ref{int.repr.1.2}) and (\ref{int.repr.1.4}) have been 
derived by using only the symmetry properties of the measure of the integral.
This implies that these properties are valid for generic values of $D$.
\section{If $N$ is odd or $N$ is even but $\gamma_\chi$ is present: the Clifford Algebra}
\label{sec:odd}
We represent the trace involving a single $\gamma_\chi$ or
an odd number of gamma's by an integral over Grassmann
variables $\{\bar c_j\}$ and $\{\chi_\mu\}$. 
\begin{eqnarray}&&
Tr(\not\! p_1\dots \not\! p_{_N}\gamma_\chi)
\nonumber\\&&  
=
i^{\frac{D(D-1)}{2}} K \int d^{^D}\chi\, d^{^N} \bar c\, 
\exp \Big(\sum_{\mu=1}^D\sum_{i=1}^N\bar c_i p_{i\mu}\chi_\mu 
+\sum_{i<j}^N\bar c_i( p_{i} p_j) \bar c_j\Big ) .
\label{int.repr.1.8}
\end{eqnarray}
%
As before the algebra of the gamma's is correctly 
implemented; in fact 
\begin{eqnarray}&&
Tr(\not\! p_2 \not\! p_1\not\! p_3\dots \not\! p_{_N}\gamma_\chi)
=
i^{\frac{D(D-1)}{2}} K \int d^{^D}\chi\, d^{^N} \bar c\, 
\exp \Big[\sum_{\mu=1}^D\Big(\bar c_1 p_{2\mu}\chi_\mu
\nonumber\\&&  
 +c_2 p_{1\mu}\chi_\mu  +
\sum_{i=3}^N\bar c_i p_{i\mu}\chi_\mu \Big)
+\bar c_1( p_{2} p_1) \bar c_2+\sum_{j=3}^N\bar c_1( p_{2} p_j) \bar c_j
\nonumber\\&&  
+\sum_{j=3}^N\bar c_2( p_{1} p_j) \bar c_j
+\sum_{i=3,j>i}^N\bar c_i( p_{i} p_j) \bar c_j\Big ].
\label{int.repr.1.8.1}
\end{eqnarray}
Again we rename $\bar c_1\leftrightarrow\bar c_2$ and 
 change the order in the measure
\begin{eqnarray}&&
Tr(\not\! p_2 \not\! p_1\not\! p_3\dots \not\! p_{_N}\gamma_\chi)
= -
i^{\frac{D(D-1)}{2}} K \int d^{^D}\chi\, d^{^N} \bar c\, 
\exp \Big[\sum_{\mu=1}^D\Big(\bar c_2 p_{2\mu}\chi_\mu 
\nonumber\\&&  
+c_1 p_{1\mu}\chi_\mu+
\sum_{i=3}^N\bar c_i p_{i\mu}\chi_\mu \Big)
+\bar c_2( p_{2} p_1) \bar c_1+\sum_{j=3}^N\bar c_2( p_{2} p_j) \bar c_j
\nonumber\\&&  
+\sum_{j=3}^N\bar c_1( p_{1} p_j) \bar c_j
+\sum_{i=3,j>i}^N\bar c_i( p_{i} p_j) \bar c_j\Big ]
\nonumber\\&&  
= -
i^{\frac{D(D-1)}{2}} K \int d^{^D}\chi\, d^{^N} \bar c\, 
\exp \Big[\sum_{\mu=1}^D\Big(
\sum_{i=1}^N\bar c_i p_{i\mu}\chi_\mu \Big)
-2\bar c_1( p_{2} p_1) \bar c_2
\nonumber\\&&  
+\sum_{j=3}^N\bar c_2( p_{2} p_j) \bar c_j
+\sum_{j=2}^N\bar c_1( p_{1} p_j) \bar c_j
+\sum_{i=3,j>i}^N\bar c_i( p_{i} p_j) \bar c_j\Big ].
\label{int.repr.1.8.2}
\end{eqnarray}
Thus one gets
\begin{eqnarray}&&
Tr(\{\not\! p_1,\not\!p_2\}\not\!p_3\dots \not\! p_{_N}\gamma_\chi)
= 
i^{\frac{D(D-1)}{2}} K \int d^{^D}\chi\, d^{^N} \bar c\, 
\exp \Big[\sum_{\mu=1}^D\Big(
\sum_{i=1}^N\bar c_i p_{i\mu}\chi_\mu \Big)\Big ]
\nonumber\\&&  
\Big(1-\exp(-2\bar c_1( p_{2} p_1) \bar c_2)\Big)
\exp \Big[
\sum_{i=1,j>i}^N\bar c_i( p_{i} p_j) \bar c_j\Big ]
\nonumber\\&&  
= 
i^{\frac{D(D-1)}{2}} K \int d^{^D}\chi\, d^{^N} \bar c\, 
\Big(2\bar c_1( p_{2} p_1) \bar c_2\Big)
\exp \Big[\sum_{\mu}^D\Big(
\sum_{i=3}^N\bar c_i p_{i\mu}\chi_\mu \Big)\Big ]
\nonumber\\&&  
\exp \Big[
\sum_{i=3,j>i}^N\bar c_i( p_{i} p_j) \bar c_j\Big ]
= 2( p_{2} p_1)
Tr(\not\! p_3\dots \not\! p_{_N}\gamma_\chi)\, ,
\label{int.repr.1.9}
\end{eqnarray}
after the integration over $d\bar c_2 d\bar c_1$.
\par
The result of eq. (\ref{int.repr.1.9}) has been obtained
for generic values of $D$, in fact the integration over $\chi$'s is not
involved in the proof. 
\section{Cyclicity of Trace for generic $D$ with $\gamma_\chi$ or odd $N$?}
\label{sec:cycl}
{If $\g5$ is present, cyclicity is a more complex issue. In fact
this property of the trace would require
\begin{eqnarray}&&
Tr(\not\! p_1\not\!p_2\not\!p_3\dots \not\! p_{_N}\gamma_\chi)
=
Tr(\not\!p_2\not\!p_3\dots \not\! p_{_N}\gamma_\chi\not\! p_1).
\label{cycl.1}
\end{eqnarray}
In the above equation the LHS is defined by eq. (\ref{int.repr.1.8}),
but the RHS has not been defined yet. In the present approach
we require that eq. (\ref{cycl.1}) \underline{defines} the RHS;
thus preserving cyclicity.
}
\par
{Let us see what are the consequences of this assumption.}
From repeated use of eq. (\ref{int.repr.1.9}) 
it follows that ($~\hat{}  $ means that the factor has to be omitted )
\begin{eqnarray}&&
Tr(\not\! p_1\not\!p_2\not\!p_3\dots \not\! p_{_N}\gamma_\chi)
= 2 \sum_{k=2}^N(-)^k(p_1p_k)
Tr(\widehat{\not\! p_1}\dots\widehat{\not\!p_k}\dots \not\! p_{_N}\gamma_\chi)
\nonumber\\&&
+(-)^{(N-1)}
Tr(\not\! p_2\dots \not\! p_{_N} \not\! p_{1} \gamma_\chi).
\label{int.repr.1.9.1}
\end{eqnarray}
Then cyclicity implies:
\begin{itemize}
\item for even $N$
\begin{eqnarray}&&
Tr(\not\!p_2\not\!p_3\dots \not\! p_{_N}{\{\gamma_\chi,\not\! p_1\}})
=
Tr(\{\not\! p_1,\not\!p_2\not\!p_3\dots \not\! p_{_N}\}{\gamma_\chi})
\nonumber\\&&
= 2 \sum_{k=2}^N(-)^k(p_1p_k)
Tr(\widehat{\not\! p_1}\dots\widehat{\not\!p_k}\dots \not\! p_{_N}
{\gamma_\chi})
\label{int.repr.1.9.2}
\end{eqnarray}
\item for odd $N$
\begin{eqnarray}&&
Tr(\not\!p_2\not\!p_3\dots \not\! p_{_N}{[\gamma_\chi,\not\! p_1]})
=
Tr([\not\! p_1,\not\!p_2\not\!p_3\dots \not\! p_{_N}]{\gamma_\chi})
\nonumber\\&&
= 2 \sum_{k=2}^N(-)^k(p_1p_k)
Tr{(}\widehat{\not\! p_1}\dots\widehat{\not\!p_k}\dots \not\! p_{_N}{\gamma_\chi}).
\label{int.repr.1.9.3}
\end{eqnarray}
\end{itemize}
It should be stressed that eqs. (\ref{int.repr.1.9.2}) and 
(\ref{int.repr.1.9.3}) have been
obtained with no conditions on $D$. Moreover it is worth noticing
that both eqs. (\ref{int.repr.1.9.2}) and  (\ref{int.repr.1.9.3})
are consistent with Lorentz covariance.
\par
The algebra of $\gamma_\chi$ will be tested in the evaluation
of the ABJ anomaly and in the validity of the LFE.
%
\section{Moving $\gamma_\chi$ around}
\label{sec:moving}

Let us elaborate on the conclusions of Section \ref{sec:cycl}
and in particular on the implications of cyclicity. 
We now demonstrate that cyclity allows us to represent a situation
where $\gamma_\chi$ is in arbitrary position.
\par
We have made the assumption: represent a trace with one
$\gamma_\chi$ to the right (eq. (\ref{int.repr.1.8})) by
\begin{eqnarray}&&
Tr(\not\! p_{1} \not\! p_2\dots \not\! p_{N}\gamma_\chi )
=
i^{\frac{D(D-1)}{2}} K \int d^{^D}\chi\, d^{^N} \bar c\, 
\exp \Big(\sum_{\mu=1}^D\sum_{i=1}^N\bar c_i  p_{i\mu}\chi_\mu 
\nonumber\\&&  
+\sum_{i=1,i<j}^N\bar c_i( p_{i} p_j) \bar c_j\Big) 
.
\label{nothappy.1}
\end{eqnarray}
With the same tools we want to represent
\begin{eqnarray}
Tr(\not\! p_{1} \not\! p_2\dots \not\! p_{N-1}\gamma_\chi \not\! p_{N})
.
\label{nothappy.2}
\end{eqnarray}
To achieve this, we consider the expression as in eq.
(\ref{nothappy.1}) but with $ \not\! p_{N}$ in the first position
\begin{eqnarray}&&
Tr( \not\! p_{N}\not\! p_{1} \not\! p_2\dots \not\! p_{N-1}\gamma_\chi )
=
i^{\frac{D(D-1)}{2}} K \int d^{^D}\chi\, d^{^N} \bar c\, 
\exp \Big(\sum_{\mu=1}^D\sum_{i=2}^{N}\bar c_i  p_{(i-1)\mu}\chi_\mu 
\nonumber\\&&  
+\sum_{\mu=1}^D\bar c_1  p_{N\mu}\chi_\mu 
+\sum_{i=2,i<j}^N\bar c_i( p_{i-1} p_{j-1}) \bar c_j
+ \sum_{j=2}^N\bar c_1 (p_N p_{j-1}) \bar c_j\Big).
\label{nothappy.3}
\end{eqnarray}
Finally we use cyclicity to obtain
\begin{eqnarray}&&
Tr( \not\! p_{1} \not\! p_2\dots \not\! p_{N-1}\gamma_\chi \not\! p_{N})
=
i^{\frac{D(D-1)}{2}} K \int d^{^D}\chi\, d^{^N} \bar c\, 
\exp \Big(\sum_{\mu=1}^D\sum_{i=2}^{N}\bar c_i  p_{(i-1)\mu}\chi_\mu 
\nonumber\\&&  
+\sum_{\mu=1}^D\bar c_1  p_{N\mu}\chi_\mu 
+\sum_{i=2,i<j}^N\bar c_i( p_{i-1} p_{j-1}) \bar c_j
+ \sum_{j=2}^N\bar c_1 (p_N p_{j-1}) \bar c_j\Big).
\label{nothappy.3.1}
\end{eqnarray}
We can write a different expression if we rename the
dummy integration variables
\begin{eqnarray}&&
\bar c_j \to \bar c_{j-1}\quad {\rm for} \quad j>1 
\nonumber\\&& 
\bar c_1 \to \bar c_N
.
\label{nothappy.4}
\end{eqnarray}
We get
\begin{eqnarray}&&
Tr( \not\! p_{1} \not\! p_2\dots \not\! p_{N-1}\gamma_\chi \not\! p_{N})
\nonumber\\&&  
= (-)^{N-1}
i^{\frac{D(D-1)}{2}} K \int d^{^D}\chi\, d^{^N} \bar c\, 
\exp \Big(\sum_{\mu=1}^D\sum_{i=2}^{N}\bar c_{i-1}  p_{(i-1)\mu}\chi_\mu 
\nonumber\\&&  
+\sum_{\mu=1}^D\bar c_N  p_{N\mu}\chi_\mu 
+\sum_{i=2,i<j}^N\bar c_{i-1}( p_{i-1} p_{j-1}) \bar c_{j-1}
+ \sum_{j=2}^N\bar c_N (p_N p_{j-1}) \bar c_{j-1}\Big)
\nonumber\\&&  
= (-)^{N-1}
i^{\frac{D(D-1)}{2}} K \int d^{^D}\chi\, d^{^N} \bar c\, 
\exp \Big(\sum_{\mu=1}^D\sum_{i=1}^{N}\bar c_{i}  p_{i\mu}\chi_\mu 
\nonumber\\&&  
+\sum_{i=1,i<j}^{N-1}\bar c_{i}( p_{i} p_{j}) \bar c_{j}
- \sum_{j=1}^{N-1}\bar c_j (p_j p_{N}) \bar c_{N}\Big)
.
\label{nothappy.3.2}
\end{eqnarray}
{\bf Comment}: eq. (\ref{nothappy.3.2}) can be written with the usual
trick as in eq. (\ref{int.repr.1.9})
\begin{eqnarray}&&
Tr( \not\! p_{1} \not\! p_2\dots \not\! p_{N-1}\gamma_\chi \not\! p_{N})
\nonumber\\&&  
= (-)^{N-1}
i^{\frac{D(D-1)}{2}} K \int d^{^D}\chi\, d^{^N} \bar c\, 
\prod_{j=1}^{N-1}\Big(1-2\bar c_j(p_j p_{N})\bar c_N\Big)
\nonumber\\&&  
\exp \Big(\sum_{\mu=1}^D\sum_{i=1}^{N}\bar c_{i}  p_{i\mu}\chi_\mu 
+\sum_{i=1,i<j}^{N-1}\bar c_{i}( p_{i} p_{j}) \bar c_{j}
+ \sum_{j=1}^{N-1}\bar c_j (p_j p_{N}) \bar c_{N}\Big)
\nonumber\\&& 
= (-)^{N-1}Tr( \not\! p_{1} \not\! p_2\dots \not\! p_{N-1} \not\! p_{N}\gamma_\chi)
\nonumber\\&&  
-2(-)^{N-1} \sum_{j=1}^{N-1}(p_j p_{N})(-)^{N-j-1}
Tr( \not\! p_{1}\dots \widehat{\not\! p_j}\dots \not\! p_{N-1} 
\widehat{\not\! p_{N}}\gamma_\chi)
.
\label{nothappy.3.2.0}
\end{eqnarray}
The last expression is an alternative definition of the shifted $\gamma_\chi$:
\begin{eqnarray}&&
Tr( \not\! p_{1} \not\! p_2\dots \not\! p_{N-1}\gamma_\chi \not\! p_{N})
\nonumber\\&&  
= (-)^{N-1}Tr( \not\! p_{1} \not\! p_2\dots \not\! p_{N-1} \not\! p_{N}\gamma_\chi)
\nonumber\\&&  
+Tr( \not\! p_{1} \not\! p_2\dots \not\! p_{N-1}\gamma_\chi \not\! p_{N})
+(-)^{N}Tr( \not\! p_{1} \not\! p_2\dots \not\! p_{N-1} \not\! p_{N}\gamma_\chi)
\label{nothappy.3.2.1}
\end{eqnarray}
For even $N$
\begin{eqnarray}&&
Tr( \not\! p_{1} \not\! p_2\dots \not\! p_{N-1}\gamma_\chi \not\! p_{N})
\nonumber\\&&  
= - Tr( \not\! p_{1} \not\! p_2\dots \not\! p_{N-1} \not\! p_{N}\gamma_\chi)
+Tr(\big\{\not\! p_{N}, \not\! p_{1} \not\! p_2\dots \not\! p_{N-1}
\big\}\gamma_\chi )
\label{nothappy.3.2.2}
\end{eqnarray}
For odd $N$
\begin{eqnarray}&&
Tr( \not\! p_{1} \not\! p_2\dots \not\! p_{N-1}\gamma_\chi \not\! p_{N})
\nonumber\\&&  
=  Tr( \not\! p_{1} \not\! p_2\dots \not\! p_{N-1} \not\! p_{N}\gamma_\chi)
+Tr(\big[\not\! p_{N}, \not\! p_{1} \not\! p_2\dots \not\! p_{N-1}
\big]\gamma_\chi ).
\label{nothappy.3.2.3}
\end{eqnarray}
Eqs. (\ref{nothappy.3.2.2}) and (\ref{nothappy.3.2.2}) are in agreement
with eqs. (\ref{int.repr.1.9.2}) and (\ref{int.repr.1.9.3}).

%
%
\section{{Unfolding $\gamma_\chi$: as a Product of Gamma's}}
\label{sec:g5}
For integer values of $D$, eq. (\ref{int.repr.1.8})
tells that the trace is zero, unless $N\ge D$,
{in fact the integration over $d^{^D}\chi$ is non-zero
only if there are as many $\bar c_j$ as $\chi_\mu$.
Moreover $N-D$ must be even}. In particular for $N=D$
\begin{eqnarray}
Tr(\not\! p_1\not\! p_2\dots \not\! p_{_D}\g5) 
=i^{\frac{D(D-1)}{2}} K \det[p]= i^{\frac{D(D-1)}{2}}
K\epsilon_{\mu_1\dots\mu_D}p_{1\mu_1}\dots p_{_D\mu_D}\, ,
\label{g5.3}
\end{eqnarray}
{
that is $\gamma_\chi$ is proportional to the product
of all gamma's. It is given by the unity for odd $D$.}
\par
This property can be traced in the integral representation
(\ref{int.repr.1.8}).
For the general case $N>D$ it is convenient to write
eq. (\ref{int.repr.1.8}) in a different fashion.
We first change notation: for $i=1,\dots,D$
\begin{eqnarray}&&
\bar c_{_N+i} \equiv \chi_i,
\nonumber\\&&
p_{(_N+i)\mu} = \delta_{i\mu}
.
\label{g5.4}
\end{eqnarray}
Then we have
\begin{eqnarray}&&
(p_i p_j) = \delta_{ij},\quad {\rm for}~ i,j>N
\nonumber\\&&
p_{i\mu}= (p_ip_{(_N+\mu)}),\quad {\rm for}~ i<N
.
\label{g5.5}
\end{eqnarray}
With the new variables,  eq. (\ref{int.repr.1.8})
becomes 
\begin{eqnarray}&&
Tr(\not\! p_1\dots \not\! p_{_N}\gamma_\chi)
\nonumber\\&&  
=
i^{\frac{D(D-1)}{2}} K \int  d\bar c_{_{(N+D)}} \dots d\bar c_1\, \,
\exp \Big(
\sum_{i=1,i<j}^{(N+D)}\bar c_i( p_{i} p_j) \bar c_j\Big ) 
\label{g5.6}
\end{eqnarray}
or
\begin{eqnarray}&&
Tr(\not\! p_1\dots \not\! p_{_N}\not\!p_{_N+1}\dots\not\!p_{_N+D})
\nonumber\\&&  
=
 K \int  d\bar c_{_{(N+D)}} \dots\bar c_1\, \, 
\exp \Big(
\sum_{i=1,i<j}^{(N+D)}\bar c_i( p_{i} p_j) \bar c_j\Big ).
\label{g5.7}
\end{eqnarray}
Once the trace is written in the canonical form eq. (\ref{g5.6})
or eq. (\ref{g5.7}) one can apply the results of Section \ref{sec:int.rep}
about the Clifford algebra in eq. (\ref{int.repr.1.2.0}) and
those of Section \ref{sec:even_cycl} for cyclicity as 
in eq. (\ref{int.repr.1.4}).
\par
This argument shows that the properties of the integral representations
are identical for integer $D$ to those of the standard irreducible 
representation in terms of matrices.
\subsection{{More over Cyclicity}}
\label{sec:cyc}
The above argument is the final answer to the check
of the integral representation for integer $D$ . However we shall explicitly
verify some of the formulae in order to get used to the relevant
algebraic manipulations.
In particular we consider again the issue
of cyclicity for integer $D$ and $N>D$. 
\par
By using cyclicity as in eq. (\ref{int.repr.1.4}) and the gamma's algebra one gets
\begin{eqnarray}
Tr(\not\! p_1\not\! p_2\dots \not\! p_{_N}\g5) 
=(-)^{(D-1)}Tr(\not\! p_2\dots \not\! p_{_N}\not\! p_1\g5)
.
\label{int.repr.1.11}
\end{eqnarray}
{
Let us give a formal proof of the above equation by using the representation
(\ref{g5.7})
\begin{eqnarray}&&
Tr(\not\! p_2\dots \not\! p_{_N}\not\! p_1\not\!p_{_{N+1}}\dots\not\!p_{_{N+D}})
\nonumber\\&&  
=
 K \int  d\bar c_{_{(N+D)}} \dots \,
d\bar c_{_N}\dots d\bar c_1 \, 
\exp \Big(
\sum_{i=1,i<j\le N}^{^{N-1}}\Big[\bar c_i( p_{i+1} p_{j+1}) \bar c_j\Big] \Big|_{p_{N+1}=p_1} 
\nonumber\\&&  
+ \sum_{i=1}^{N}\sum_{j=1}^D\bar c_i (p_{i+1}\Big|_{p_{N+1}=p_1}  p_{N+j})\bar c_{N+j}
\Big)
\label{g5.8}
\end{eqnarray}
and a similar one for
\begin{eqnarray}&&
Tr(\not\! p_2\dots \not\! p_{_N}\not\!p_{_{N+1}}\dots\not\!p_{_{N+D}}\not\! p_1)
\nonumber\\&&  
=
 K \int  d\bar c_{_{(N+D)}} \dots \,
d\bar c_{_N}\dots d\bar c_1 \, 
\exp \Big(
\sum_{i=1,i<j< N}^{^N}\Big[\bar c_i( p_{i+1} p_{j+1}) \bar c_j\Big] 
\nonumber\\&&  
+ \sum_{i=1}^{N-1}
\bar c_i (p_{i+1}  p_{1})\bar c_{N+D}+
 \sum_{i=1}^{N-1}\sum_{j=1}^D\bar c_i (p_{i+1}  p_{N+j})\bar c_{N+j-1}
\Big).
\label{g5.9}
\end{eqnarray}
We need 
\begin{eqnarray}&&
Tr(\not\! p_2\dots \not\! p_{_N}\not\! p_1\not\!p_{_{N+1}}\dots\not\!p_{_{N+D}})
+(-)^{D}
Tr(\not\! p_2\dots \not\! p_{_N}\not\!p_{_{N+1}}\dots\not\!p_{_{N+D}}\not\! p_1)
\nonumber\\&&  
=
 K \int  d\bar c_{_{(N+D)}} \dots \,
d\bar c_{_N}\dots d\bar c_1 \, 
\exp \Big(
\sum_{i=1,i<j\le N}^{^{N-1}}\Big[\bar c_i( p_{i+1} p_{j+1}) \bar c_j\Big] \Big|_{p_{N+1}=p_1} 
\nonumber\\&&  
+ \sum_{i=1}^{N}\sum_{j=1}^D\bar c_i (p_{i+1}\Big|_{p_{N+1}=p_1}  p_{_{N+j}})\bar c_{_{N+j}}
\Big)
\nonumber\\&&  
+(-)^{D}
 K \int  d\bar c_{_{(N+D)}} \dots \,
d\bar c_{_N}\dots d\bar c_1 \, 
\exp \Big(
\sum_{i=1,i<j< N}^{^N}\Big[\bar c_i( p_{i+1} p_{j+1}) \bar c_j\Big] 
\nonumber\\&&  
+ \sum_{i=1}^{N+D-1}
\bar c_i (p_{i+1}  p_{1})\bar c_{_{N+D}} + \sum_{i=1}^{N-1}\sum_{j=1}^{^D}
\bar c_i (p_{i+1}  p_{_{N+j}})\bar c_{_{N+j-1}}
\Big).
\label{g5.10}
\end{eqnarray}
In the second term in eq. (\ref{g5.10}) we rename $\bar c$ 
according to the following table
\begin{eqnarray}&&
\bar c_{{N}}  \to \bar c_{{N+1}}
\nonumber\\&&   
\bar c_{{N+j}}  \to \bar c_{{N+j+1}}, \quad j<D
\nonumber\\&& 
\bar c_{_{N+D}} \to \bar c_{_{N}}
 .
\label{g5.11}
\end{eqnarray}
Next,  we recover the order of the product in the measure.
Thus we get a factor $(-)^D$
\begin{eqnarray}&&
Tr(\not\! p_2\dots \not\! p_{_N}\not\! p_1\not\!p_{_{N+1}}\dots\not\!p_{_{N+D}})
+(-)^{D}
Tr(\not\! p_2\dots \not\! p_{_N}\not\!p_{_{N+1}}\dots\not\!p_{_N+D}\not\! p_1)
\nonumber\\&&  
=
 K \int  d\bar c_{_{(N+D)}} \dots \,
d\bar c_{_N}\dots d\bar c_1 \, \Bigg\{
\exp \Big(
\sum_{i=1,i<j\le N}^{^{N-1}}\Big[\bar c_i( p_{i+1} p_{j+1}) \bar c_j\Big] \Big|_{p_{N+1}=p_1} 
\nonumber\\&&  
+ \sum_{i=1}^{N}\sum_{j=1}^D\bar c_i (p_{i+1}\Big|_{p_{N+1}=p_1}  p_{_{N+j}})\bar c_{N+j}
\Big)
\nonumber\\&&  
+
\exp \Big(
\sum_{i=1,i<j< N}^{^N}\Big[\bar c_i( p_{i+1} p_{j+1}) \bar c_j\Big] 
+ \sum_{i=1}^{N-1}\bar c_i (p_{i+1}  p_{1})\bar c_{_{N}}
\nonumber\\&&   
-\sum_{i=1}^{D}\bar c_{_{N}}  ( p_{1} p_{_{N+i}} )\bar c_{_{N+i}}
+ \sum_{i=1}^{N-1}\sum_{j=1}^{^D}
\bar c_i (p_{i+1}  p_{_{N+j}})\bar c_{_{N+j}}
\Big)\Bigg\}
\label{g5.12}
\end{eqnarray}
%
\begin{eqnarray}
&&  
=
 K \int  d\bar c_{_{(N+D)}} \dots \,
d\bar c_{_N}\dots d\bar c_1 \, \Big(1+
\prod_{i=1}^D\Big[1-2\bar c_{_{N}}  ( p_{1} p_{_{N+i}} )\bar c_{_{N+i}} \Big]\Big) 
\nonumber\\&&  
\exp \Big\{
\sum_{i=1,i<j\le N}^{^{N-1}}\Big[\bar c_i( p_{i+1} p_{j+1}) \bar c_j\Big] \Big|_{p_{N+1}=p_1}
+ \sum_{i=1}^{N}\sum_{j=1}^D\bar c_i (p_{i+1}\Big|_{p_{N+1}=p_1}
\nonumber\\&&  
  p_{_{N+j}})\bar c_{N+j}
\Big\}
.
\label{g5.13}
\end{eqnarray}
The result can be easily interpreted as
\begin{eqnarray}&&
{\rm eq.} (\ref{g5.10})
= 2Tr\Big(\not\! p_2\dots \not\! p_{_N}\not\! p_1\not\!p_{_{N+1}}\dots\not\!p_{_{N+D}}\Big)
\nonumber\\&&  
-2\sum_{j=1}^D(p_1, p_{_{N+j}})(-)^{j-1}
Tr\Big(\not\! p_2\dots \not\! p_{_N}\widehat{\not\! p_1}\not\!p_{_{N+1}}\dots
\widehat{\not\!p_{_{N+j}}}\dots \not\!p_{_{N+D}}\Big
).
\label{g5.14}
\end{eqnarray}
For even $D$ eq. (\ref{g5.14}) becomes
\begin{eqnarray}&&
 Tr\Big(\not\! p_2\dots \not\! p_{_N}\Big[ \not\! p_1,
\not\!p_{_{N+1}}\dots\not\!p_{_{N+D}}\big]\Big)
\nonumber\\&&  
=2\sum_{j=1}^D(p_1, p_{_{N+j}})(-)^{j-1}
Tr\Big(\not\! p_2\dots \not\! p_{_N}\widehat{\not\! p_1}\not\!p_{_{N+1}}\dots
\widehat{\not\!p_{_{N+j}}}\dots \not\!p_{_{N+D}}
\Big).
\label{g5.15}
\end{eqnarray}
and for odd $D$
\begin{eqnarray}&&
 Tr\Big(\not\! p_2\dots \not\! p_{_N}\Big\{ \not\! p_1,
\not\!p_{_{N+1}}\dots\not\!p_{_{N+D}}\Big\}\Big)
\nonumber\\&&  
=2\sum_{j=1}^D(p_1, p_{_{N+j}})(-)^{j-1}
Tr\Big(\not\! p_2\dots \not\! p_{_N}\widehat{\not\! p_1}\not\!p_{_{N+1}}\dots
\widehat{\not\!p_{_{N+j}}}\dots \not\!p_{_{N+D}}
\Big).
\label{g5.16}
\end{eqnarray}
Finally eq. (\ref{int.repr.1.11}) is obtained from eqs. (\ref{g5.15}) and 
(\ref{g5.16}) by using the identity 
\begin{eqnarray}&&
2\sum_{j=1}^D(p_1, p_{_{N+j}})(-)^{j-1}
\not\!p_{_{N+1}}\dots
\widehat{\not\!p_{_{N+j}}}\dots \not\!p_{_{N+D}}
\nonumber\\&&  
=
2\sum_{j=1}^D(p_1)_j(-)^{j-1}
\not\!p_{_{N+1}}\dots
\widehat{\not\!p_{_{N+j}}}\dots \not\!p_{_{N+D}}
\nonumber\\&&  
=
2\sum_{j=1}^D(p_1)_j(-)^{j-1}
\not\!p_{_{N+1}}\dots
\widehat{\not\!p_{_{N+j}}}\gamma_j^2\dots \not\!p_{_{N+D}}
\nonumber\\&&  
=
2\not \! p_1
\not\!p_{_{N+1}}\dots
 \not\!p_{_{N+D}}.
\label{g5.17}
\end{eqnarray}

}
The  relation  (\ref{int.repr.1.11}) implies
\begin{eqnarray}
\left \{
\begin{array}{l}
\{\g5,\gamma_\nu\} =0,\quad \forall \nu=1\dots D \quad {\rm for~even~} D \\
\   [\g5, \gamma_\nu]  =0,\quad \forall \nu=1\dots D \quad {\rm for~odd~} D\, .
\end{array}
\right.
\label{int.repr.1.12}
\end{eqnarray}
We match the representation of the trace in eq. (\ref{int.repr.1.8})
with the standard matrix expressions.
The matrix representation of the gamma's is assumed to be irreducible thus
we choose the  phase
\begin{eqnarray}&&
\gamma_\chi= (i)^{\frac{D(D-1)}{2}}  \gamma_1\dots \gamma_{_D}\quad {\rm for~
even}~ D 
\nonumber\\&&  
\gamma_\chi= (i)^{\frac{D(D-1)}{2}}\mathbb{I} \quad {\rm for~ odd}~ D .
\label{int.repr.1.8.0}
\end{eqnarray}
%


\section{When $D$ is an Integer: Trace in Closed Form}
\label{sec:integ}
Now we study the integral representation in eq. (\ref{int.repr.1.8}) for
integer $D$. 
By using a theorem
in general matrix theory \cite{golub2013}, the rectangular matrix $p_{j\mu}, \,\,
\scriptstyle{j=1,\dots,N}$ 
can be brought to a diagonal form $\Sigma$ by suitable  
unitary transformations  $  U$ and  $V^\dagger$ 
\begin{eqnarray}
p=  U \Sigma V^\dagger.
\label{int.repr.1.10}
\end{eqnarray}
$U$ is a $N\times N$ and  $V$ is $D\times D$  matrix (singular value
decomposition).
Both $U$ and $V$ can be orthogonal matrices if $p$ is a real matrix.
$\Sigma$ is unique if the eigenvalues are positive or zero and ordered
\begin{eqnarray}
\Sigma_{i\mu} = \sigma_i \delta_{i\mu}, \qquad \sigma_i \ge \sigma_{i+1}\ge 0,
\qquad 1\le i< min(D,N).
\label{int.repr.1.10.-1}
\end{eqnarray}
We can change variable of integration by  unitary transformation
\begin{eqnarray}&&
\bar c_i \to \bar c_j U^\dagger_{ji}
\nonumber \\&&
\chi_\mu\to V_{\mu\mu'}  \chi_{\mu'} \, .
\label{int.repr.1.10.0.0}
\end{eqnarray}
The ensuing integrations on the $\chi$'s  
can be non-zero only if the rank of the matrix $\{p\}$
is  equal to $D$. Thus it is necessary that
$N\ge D$. Consequently if $N>D$ the integration over $\bar c_j$
is non zero only if $N-D$ is even. Thus
\begin{eqnarray}
n=\frac{1}{2}(N-D) 
\label{int.repr.1.10.0.-3}
\end{eqnarray}
is a positive integer.
\par
The bilinear term in $\bar c$ can be written (as in eq. (\ref{int.repr.1.0}))
\begin{eqnarray}&&
\sum_{i,j=1,i<j}^N \bar c_i (p_i p_j) \bar c_j =
\frac{1}{2} \sum_{i,j}\bar c_i h_{[ij]}\bar c_j, 
\nonumber \\&&
h_{ij}|_{i<j}= (p_i p_j) = U_{ii'}\sigma^2_{i'} U^\dagger_{i'j}.
\label{int.repr.1.10.0.-2}
\end{eqnarray}
The integral representation (\ref{int.repr.1.8}) becomes
\begin{eqnarray}&&
Tr(\not\! p_1\dots \not\! p_{_N}\gamma_\chi)
=
i^{\frac{D(D-1)}{2}} K \int d^{^D}\chi\, d^{^N} \bar c\, 
\nonumber\\&&  
\exp \Big(\sum_{\mu=1}^D\sum_{i=1}^D\bar c_i \Sigma_{i \mu}\chi_\mu  
+\frac{1}{2}\sum_{ab=D+1}^N\sum^N_{i,j=1}\bar c_{a}U^\dagger_{ai}h_{ij}U_{jb} \bar c_b
\Big)\, ,
\label{int.repr.1.10.0.1}
\end{eqnarray}
where the limits on the sums are dictated by the form of 
$\Sigma$ as in eq. (\ref{int.repr.1.10.-1}).
\par
In the present problem both $U$ and $V$ can be chosen to be orthogonal matrices,
then the matrix 
\begin{eqnarray}
H_{ab}\equiv \sum^N_{i,j=1}U^\dagger_{ai}h_{ij}U_{jb}, \quad
a,b=D+1,\dots, N
\label{int.repr.1.10.0.2}
\end{eqnarray}
is skew-symmetric. The integration over $\bar c$ is a standard
result in matrix theory (Pfaffian) \cite{E.R.Caianiello:1952ww}
\begin{eqnarray}&&
Pf(H)=
\int \, d^{^(N-D)} \bar c\, 
\,\,\,\,   \exp
\frac{1}{2} \Big(\sum_{a,b=D+1}^N\bar c_{a}H_{ab}\bar c_b
\Big)
\nonumber\\&&  
= \sum_{\mathbb P}\, ' \delta_{\mathbb P} H{i_1j_1} H{i_2j_2}H{i_3j_3}\dots 
\label{int.repr.1.10.0.3}
\end{eqnarray}
where $\mathbb P$ is any permutation of $D+1,\dots,N$ and $\delta_{\mathbb P}$ is its signature. The sum $\sum'$ is restricted to the permutations
satisfying the conditions
\begin{eqnarray}
i_k<i_{k+1} ~{\rm and}~ i_k<j_k, \forall k\, .
\end{eqnarray}
The above expression can be evaluated by using the relation
(Thomas Muir)
\begin{eqnarray}
\Big[Pf(H)\Big]^2
= det [H ]
\, .
\label{int.repr.1.10.0.4}
\end{eqnarray}
Thus the singular value decomposition of the matrix $p_{i\mu}$
allows a straightforward evaluation of the trace in even dimension
($\gamma_\chi$ present) or in odd dimensions. 
\par
\section{ABJ Anomaly}
\label{sec:ano}
We use the algebra for $\gamma_\chi$ developed in Section \ref{sec:cycl}
in order to evaluate the ABJ anomaly \cite{Adler:1969gk}\cite{Bell:1969ts}.
We consider a massless fermion triangle, where one vertex
is given by an axial current. Thus we consider the integral
($p$ is the incoming momentum on the vertex $\sigma$ and $k$ on $\rho$; crossed
graph will be considered at the end)
\begin{eqnarray}
T_{\mu\rho\sigma}(k,p)={-i}
\int \frac{d^D q}{(2\pi)^D}~
\frac{
 Tr ~\Bigl\{\gamma_\mu\gamma_\chi~
(q-k)_\alpha\gamma_\alpha  ~\gamma_\rho ~q_\beta\gamma_\beta~ \gamma_\sigma~
(q+p)_\iota\gamma_\iota
\Bigr\}
}
{
(q-k)^2 q^2 (q+p)^2
}
\label{ano.1}
\end{eqnarray}
Now we use Feynman parametrization
and get
\begin{eqnarray}&&
 {-i}Tr ~\Bigl(\gamma_\mu\gamma_\chi~
\gamma_\alpha ~\gamma_\rho \gamma_\beta~ \gamma_\sigma~
\gamma_\iota
\Bigr)
  2 
\int_0^1 dx\int_0^x dy \int \frac{d^D q}{(2\pi)^D}~
\nonumber\\&&
(q+r-k)_\alpha~(q+r)_\beta
(q+r+p)_\iota
\nonumber\\&&
\Bigl[
q^2+k^2y+p^2 x-p^2 y-(ky-px+py)^2
\Bigr]^{-3},
\label{ano.4}
\end{eqnarray}
\begin{eqnarray}
r_\nu \equiv (yk-xp+yp)_\nu.
\label{ano.2}
\end{eqnarray}
We use the simplified case 
\begin{eqnarray}
k^2=p^2=0.
\label{ano.3}
\end{eqnarray}
After symmetric integration over $q$ we can split the integral
into a divergent
\begin{eqnarray}&&
{-i}
 \frac{2}{D}
\int_0^1 dx\int_0^x dy \Bigl(\delta_{\alpha\beta}
(r+p)_\iota +\delta_{\alpha\iota}r_\beta
+\delta_{\beta\iota}(r-k)_\alpha\Bigr)
\nonumber\\&&
\Big(-\frac{i}{(4\pi)^2} \Big)\Big[\frac{2}{D-4}+\gamma +2
-\ln 4\pi+\ln 2pky(y-x)\Big]
\label{ano.5}
\end{eqnarray}
and finite part
\begin{eqnarray}&&{-i}
 2
\int_0^1 dx\int_0^x dy (r-k)_\alpha r_\beta
(r+p)_\iota
\int \frac{d^Dq}{(2\pi)^D}\frac{1}{(q^2-2kp y(y-x))^3}
\nonumber\\&&
=-\frac{{1}}{(4\pi)^2} \frac{1}{pk}
\int_0^1 dx\int_0^x dy (r-k)_\alpha r_\beta
(r+p)_\iota\frac{1}{{2} y(y-x)}.
\label{ano.6}
\end{eqnarray}
In front of the two amplitudes (\ref{ano.5}) and (\ref{ano.6})
the gamma's trace must be expanded in powers of $(D-4)$
as required by eq. (\ref{ano.1}). For the finite part in eq.
(\ref{ano.6}) we can use the $D=4$ expression, but for the
divergent part one needs also the linear part.
Let us use the representation of $\gamma_\chi$ provided
by eq. (\ref{int.repr.1.8}) in order to tackle the problem. 
According to the discussion of the 
Section \ref{sec:cycl} we know that the algebra
of $\gamma_\chi$ for non-integer $D$ is rather complicated
as shown in eqs. (\ref{int.repr.1.9.2}) and (\ref{int.repr.1.9.3}).
Then we use instead the algebra of the gamma's in eq. (\ref{int.repr.1.2})
which has been proved valid for generic $D$; i.e. we do not
change the relative position of $\g5$  with respect to the remaining
factors in the trace. Thus in eq. (\ref{ano.1}), according to eq. (\ref{ano.5}) 
we evaluate
\begin{eqnarray}&&
Tr \Bigl(\gamma_\mu\gamma_\chi\gamma_\alpha\gamma_\rho
\gamma_\beta\gamma_\sigma\gamma_\iota \Bigr) \delta_{\alpha\beta}
=[-2 -(D-4)] Tr \Bigl(\gamma_\mu\gamma_\chi\gamma_\rho
\gamma_\sigma\gamma_\iota \Bigr)
\nonumber\\&&
Tr \Bigl(\gamma_\mu\gamma_\chi\gamma_\alpha\gamma_\rho
\gamma_\beta\gamma_\sigma\gamma_\iota \Bigr) \delta_{\beta\iota}
=[-2 -(D-4)] Tr \Bigl(\gamma_\mu\gamma_\chi\gamma_\alpha\gamma_\rho
\gamma_\sigma \Bigr)
\nonumber\\&&
Tr \Bigl(\gamma_\mu\gamma_\chi\gamma_\alpha\gamma_\rho
\gamma_\beta\gamma_\sigma\gamma_\iota \Bigr) \delta_{\alpha\iota}
= Tr \Bigl(\gamma_\mu\gamma_\chi
\nonumber\\&&
\Big[(2 -(D-4))\gamma_\rho\gamma_\beta\gamma_\sigma
-4 (\delta_{\rho\beta}\gamma_\sigma-\delta_{\rho\sigma}\gamma_\beta+\delta_{\sigma\beta}\gamma_\rho)  \Big]
 \Bigr).
\label{ano.7}
\end{eqnarray}
We collect the non-zero part associated to the amplitude (\ref{ano.5}).

 {
\subsection{Contribution of the Divergent Integral}
Start with eq. (\ref{ano.5})
\begin{eqnarray}&& {-i}
 \frac{2}{D}
\int_0^1 dx\int_0^x dy
Tr \Bigl(\gamma_\mu\gamma_\chi\gamma_\alpha\gamma_\rho
\gamma_\beta\gamma_\sigma\gamma_\iota \Bigr)
 \Bigl(\delta_{\alpha\beta}
(r+p)_\iota +\delta_{\alpha\iota}r_\beta
+\delta_{\beta\iota}(r-k)_\alpha\Bigr)
\nonumber\\&&
\Big(-\frac{i}{(4\pi)^2} \Big)\Big[\frac{2}{D-4}+\gamma +2
-\ln 4\pi+\ln 2pky(y-x)\Big]
\nonumber\\&&
=\frac{{1}}{(4\pi)^2} \frac{2}{D}
Tr \Bigl(\gamma_\mu\gamma_\chi\gamma_\rho
\gamma_\sigma\gamma_\iota \Bigr)
\int_0^1 dx\int_0^x dy 
\nonumber\\&&
\Bigl(2(3r+p-k) +(D-4)(r+p-k)\Bigr) _\iota
\nonumber\\&&
\Big[\frac{2}{D-4}+\gamma +2
-\ln 4\pi+\ln 2pky(y-x)\Big].
\label{ano.8}
\end{eqnarray}
By using
\begin{eqnarray}&&
\int_0^1dx\int_0^x dy \Big[3( yk-xp+yp)+p-k\Big]=0
\nonumber\\&&
\int_0^1dx\int_0^x dy(yk-xp+yp+p-k)
= \frac{1}{3}(p-k)
\label{ano.9}
\end{eqnarray}
one gets
\begin{eqnarray}&&{-i}
\frac{i}{(4\pi)^2} \frac{2}{D}
Tr \Bigl(\gamma_\mu\gamma_\chi\gamma_\rho
\gamma_\sigma\gamma_\iota \Bigr)
\int_0^1 dx\int_0^x dy 
\nonumber\\&&
\Bigl(2(3r+p-k) +(D-4)(r+p-k)\Bigr) _\iota
\nonumber\\&&
\Big[\frac{2}{D-4}+\gamma +2
-\ln 4\pi+\ln 2pky(x-y)\Big]
\nonumber\\&&
=\frac{{1}}{(4\pi)^2} \frac{1}{2}
Tr \Bigl(\gamma_\mu\gamma_\chi\gamma_\rho
\gamma_\sigma\gamma_\iota \Bigr)
\nonumber\\&&
\Biggl(
\int_0^1 dx\int_0^x dy
2(3r+p-k)_\iota\ln y(x-y) 
+ \frac{2}{3}(p-k)_\iota\Biggr).
\label{ano.10}
\end{eqnarray}
Finally we use
\begin{eqnarray}&&
\int_0^1dx\int_0^x dy \ln [y(x-y)] =-  \frac{3}{2}
\nonumber\\&&
\int_0^1dx\int_0^x dy~y~ \ln [y(x-y)] =-  \frac{4}{9}
\nonumber\\&&
\int_0^1dx~x~\int_0^x dy~ \ln [y(x-y)] =-  \frac{8}{9}
\label{ano.11}
\end{eqnarray}
and get
\begin{eqnarray}&&
{-i}
\frac{i}{(4\pi)^2} \frac{1}{2}
Tr \Bigl(\gamma_\mu\gamma_\chi\gamma_\rho
\gamma_\sigma\gamma_\iota \Bigr)
\nonumber\\&&
\Biggl(
\int_0^1 dx\int_0^x dy
2(3r+p-k)_\iota\ln y(x-y) 
+ \frac{2}{3}(p-k)_\iota\Biggr)
\nonumber\\&&
=\frac{{1}}{(4\pi)^2} \frac{1}{2}
Tr \Bigl(\gamma_\mu\gamma_\chi\gamma_\rho
\gamma_\sigma\gamma_\iota \Bigr)
\Biggl(-\frac{1}{3}(p-k)_\iota
+ \frac{2}{3}(p-k)_\iota\Biggr)
\nonumber\\&&
=\frac{{1}}{(4\pi)^2} \frac{1}{2}
Tr \Bigl(\gamma_\mu\gamma_\chi\gamma_\rho
\gamma_\sigma\gamma_\iota \Bigr) \frac{1}{3}(p-k)_\iota.
\label{ano.12}
\end{eqnarray}
The divergence of the axial current is obtained
by multiplying with $(p+k)^\mu$
\begin{eqnarray}&&{-i}
(k+p)^\mu\frac{i}{(4\pi)^2} \frac{1}{2}
Tr \Bigl(\gamma_\mu\gamma_\chi\gamma_\rho
\gamma_\sigma\gamma_\iota \Bigr) \frac{1}{3}(p-k)_\iota
\nonumber\\&&
=-\frac{{1}}{(4\pi)^2}
Tr \Bigl(\gamma_\chi\gamma_\rho\not\! p 
\gamma_\sigma\not\! k \Bigr) \frac{1}{3}.
\label{ano.13}
\end{eqnarray}
By adding the crossed graph
\begin{eqnarray}
(k+p)^\mu\Big(T_{\mu\rho\sigma}^{\rm DIV}(k,p) + T_{\mu\sigma\rho}^{\rm DIV}(p,k)
\Big)
=- \frac{2}{3}\frac{{1}}{(4\pi)^2}
Tr \Bigl(\gamma_\chi\gamma_\rho\not\! p 
\gamma_\sigma\not\! k \Bigr).
\label{ano.14}
\end{eqnarray}
%
\subsection{Contribution of the Convergent Integral}
Further finite terms can be evaluated directly at $D=4$. 
From eq. (\ref{ano.4}) the finite integral contribution 
to the triangular graph is (always in the case $k^2=p^2=0$)
\begin{eqnarray}&& 
T_{\mu\rho\sigma}^{\rm FIN}(k,p)
\nonumber\\&&
={-i}
 Tr ~\Bigl\{\gamma_\mu\gamma_5~
\gamma_\alpha ~\gamma_\rho \gamma_\beta~ \gamma_\sigma~
\gamma_\iota
\Bigr\}
 2 \int_0^1 dx\int_0^x dy
(k(y-1)+p(y-x))_\alpha~
\nonumber\\&&
(ky+p(y-x))_\beta
(ky+p(y-x+1))_\iota
 (-\frac{i}{(4\pi)^2})\frac{1}{4pk y(y-x)}
\label{post.11}
\end{eqnarray}
By repeated use of the identity
\begin{eqnarray}
\not\! k \gamma_\rho \not\!k = - k^2 \gamma_\rho +2k_\rho \not\!k.
\label{post.12}
\end{eqnarray}
one shows that only two forms 
\begin{eqnarray}&&
\nonumber\\&& 
Tr ~\Bigl\{\gamma_\mu\gamma_5~
\gamma_\alpha ~\gamma_\rho \gamma_\beta~ \gamma_\sigma~
\gamma_\iota
\Bigr\}p^\alpha k^\beta p^\iota =-2p_\mu
 Tr ~\Bigl\{\gamma_5~\gamma_\rho
\not\!k~ \gamma_\sigma~\not\!p\Bigr\}
\nonumber\\&& 
Tr ~\Bigl\{\gamma_\mu\gamma_5~
\gamma_\alpha ~\gamma_\rho \gamma_\beta~ \gamma_\sigma~
\gamma_\iota
\Bigr\}k^\alpha p^\beta k^\iota =-2k_\mu
 Tr ~\Bigl\{\gamma_5~\gamma_\rho
\not\!p~ \gamma_\sigma~\not\!k\Bigr\}
\label{post.13}
\end{eqnarray}
give non-zero contribution to the divergence of the current
\begin{eqnarray}&&
(k+p)^\mu
T_{\mu\rho\sigma}^{\rm FIN}(k,p)
\nonumber\\&&
={-i}
 Tr ~\Bigl\{(k+p)^\mu\gamma_\mu \gamma_5~
\gamma_\alpha ~\gamma_\rho \gamma_\beta~ \gamma_\sigma~
\gamma_\iota
\Bigr\}
 2 \int_0^1 dx\int_0^x dy
(k(y-1)+p(y-x))_\alpha~
\nonumber\\&&
(ky+p(y-x))_\beta
(ky+p(y-x+1))_\iota
 (-\frac{i}{(4\pi)^2})\frac{1}{4pk y(y-x)}
\nonumber\\&&
=\frac{{1}}{(4\pi)^2}Tr ~\Bigl\{\gamma_5~
 ~\gamma_\rho \not\! p~ \gamma_\sigma~
\not\! k \Bigr\}\int_0^1 dx\int_0^x dy(
(y-1)-(y-x+1))
\nonumber\\&&
=-\frac{{1}}{(4\pi)^2}Tr ~\Bigl\{\gamma_5~
 ~\gamma_\rho \not\! p~ \gamma_\sigma~
\not\! k \Bigr\}\frac{2}{3}.
\label{post.14}
\end{eqnarray}
The contribution of the finite integral to the divergence
is then
\begin{eqnarray}
(k+p)^\mu(T_{\mu\rho\sigma}^{\rm FIN}(k,p)+
T_{\mu\sigma\rho}^{\rm FIN}(p,k))
=
-\frac{{1}}{(4\pi)^2 } Tr ~\Bigl\{\gamma_5~\gamma_\rho
\not\!p~ \gamma_\sigma~\not\!k\Bigr\}\frac{4}{3}
\label{post.15}
\end{eqnarray}
Finally the sum of the contributions in eq. (\ref{ano.14}) and
 (\ref{post.15}) is
\begin{eqnarray}
(k+p)^\mu(T_{\mu\rho\sigma}(k,p)+
T_{\mu\sigma\rho}(p,k))
=
-\frac{ {2}}{(4\pi)^2 } Tr ~\Bigl\{\gamma_5~\gamma_\rho
\not\!p~ \gamma_\sigma~\not\!k\Bigr\},
\label{post.16}
\end{eqnarray}
which agrees with the ABJ anomaly.
}
\section{Local Functional Equation }
\label{sec:LFE}
Once we have discovered that $\gamma_\chi$ has a complicated
behavior in $D$ dimension, we must test our formalism in
the path integral. The LFE has been discussed at length in Ref.
\cite{Ferrari:2005ii}. Here we give the essential steps.
 The functional is
\begin{eqnarray}
Z[A]=\int \prod_x
\prod_{\mu} d\bar \psi_\mu(x) \prod_{\mu'} d\psi_{\mu'}(x)
e^{i{\cal S}[A]}
\label{LFE.0}
\end{eqnarray}
where the action ($e=1$) is function of the external vector
field $A_\mu(x)$
\begin{eqnarray}
{\cal S}= \int d^Dx \bar\psi(i \not \!\partial- \not\!\! A)\psi.
\label{LFE.0.1}
\end{eqnarray}
The path integral measure is Lorentz invariant. Moreover
it is invariant under the $U(1)$ local chiral transformations
\begin{eqnarray}&&
\psi \to e^{i\alpha(x)\gamma_\chi} \psi
\nonumber\\&&
\psi^\dagger \to \psi^\dagger e^{-i\alpha(x)\gamma_\chi} 
\label{LFE.1}
\end{eqnarray}
since the Jacobian of the transformation is equal one.
In fact
\begin{eqnarray}&&
\prod_{\mu} d \psi_\mu \to
 {\rm det}( e^{i\alpha(x)\gamma_\chi}) \prod_{\mu} d \psi_\mu
=  e^{i\alpha(x)Tr(\gamma_\chi)} \prod_{\mu} d \psi_\mu .
\label{LFE.5}
\end{eqnarray}
\par
Thus if we perform a substitution in the path integral
variables according to eq. (\ref{LFE.1}) the
functional $Z$ does not change. For infinitesimal parameter
$\alpha$ one gets
\begin{eqnarray}
\Big\langle \Big(-\bar\psi \gamma_0\gamma_\chi\gamma_0 
 (i \not \!\partial- \not\!\! A)\psi
+ \bar\psi 
 (i \not \!\partial- \not\!\! A)\gamma_\chi\psi
- i\partial^\mu(\bar\psi  \gamma_\mu \gamma_\chi\psi)
\Big)\Big\rangle=0,
\label{LFE.7}
\end{eqnarray}
where the brackets $\langle\cdots \rangle$ denote the mean
value with the path integral measure of eq. (\ref{LFE.0}). 
\par
If one uses the naive commutation relations of $\gamma_\chi$
(i.e. $\{\gamma_\chi,\gamma_\mu\}=0$) the first two first terms
from the left in eq. (\ref{LFE.7}) cancel out.
\par
In Section \ref{sec:int.rep} we have found that $\gamma_\chi$ 
has complicated behavior. Then one must evaluate at one loop the expressions
in eq.  (\ref{LFE.7}) according the rules of 
eqs. (\ref{int.repr.1.9.2}) and (\ref{int.repr.1.9.3}).
The results will be compared  with eq. (\ref{post.16}).
\par
A single interaction insertion gives zero since it
depends only on $k$ or $p$; never on both. Consequently
no completely antisymmetric tensor can emerge. We need two
insertions: the triangular graph. We consider only the one
that can provide some non-zero contributions
\begin{eqnarray}
T_{\rho\sigma}(k,p)=
\int \frac{d^D q}{(2\pi)^D}~
\frac{
 Tr ~\Bigl[
\cancel{(q-k)}~\gamma_\chi ~\cancel{(q-k)}~\gamma_\rho ~\cancel{q}~ \gamma_\sigma~
\cancel{(q+p)} \Bigr]
}
{
(q-k)^2 q^2 (q+p)^2
}.
\label{LFE.8}
\end{eqnarray}
The gamma's algebra gives
\begin{eqnarray}&& 
Tr ~\Bigl[
\cancel{(q-k)}~\gamma_\chi ~\cancel{(q-k)}~\gamma_\rho ~\cancel{q}~ \gamma_\sigma~
\cancel{(q+p)} \Bigr]
\nonumber\\&& 
= -
Tr ~\Bigl[\gamma_\chi ~
\cancel{(q-k)}~\cancel{(q-k)}~\gamma_\rho ~\cancel{q}~ \gamma_\sigma~
\cancel{(q+p)} \Bigr]
\nonumber\\&&
+ Tr ~\Bigl[\gamma_\chi ~\Big\{
\cancel{(q-k)},~\cancel{(q-k)}~\gamma_\rho ~\cancel{q}~ \gamma_\sigma~
\cancel{(q+p)}\Big\} \Bigr]
\label{LFE.9}
\end{eqnarray}
In the first term of the RHS the dependence on $k$ disappears, thus it
can be neglected. We consider the remaining terms
%
\begin{eqnarray}&& 
Tr ~\Bigl[\gamma_\chi ~\Big\{
\cancel{(q-k)},~\cancel{(q-k)}~\gamma_\rho ~\cancel{q}~ \gamma_\sigma~
\cancel{(q+p)}\Big\} \Bigr]
\nonumber\\&&
= 2 (q-k)^2 Tr ~\Bigl[\gamma_\chi ~\gamma_\rho ~\cancel{q}~ \gamma_\sigma~
\cancel{(q+p)}
\Bigr]
\nonumber\\&&
 -2 (q-k)_\rho
Tr ~\Bigl[\gamma_\chi ~ 
\cancel{(q-k)} ~\cancel{q}~ \gamma_\sigma~
\cancel{(q+p)} 
\Bigr]
\nonumber\\&&
+ 2 (q-k)q
Tr ~\Bigl[\gamma_\chi ~ 
\cancel{(q-k)}~\gamma_\rho ~ \gamma_\sigma~
\cancel{(q+p)} 
\Bigr]
\nonumber\\&&
 -2 (q-k)_\sigma
Tr ~\Bigl[\gamma_\chi ~ 
\cancel{(q-k)}~\gamma_\rho ~\cancel{q}~
\cancel{(q+p)} 
\Bigr]
\nonumber\\&&
+ 2 (q-k)(q+p)
Tr ~\Bigl[\gamma_\chi ~ 
\cancel{(q-k)}~\gamma_\rho ~ \cancel{q} \gamma_\sigma
\Bigr]
\label{LFE.10}
\end{eqnarray}
%
%
All terms containing $(\not\!q)^2$ or $(~\cancel{(q-k)}~)^2$
or $(~\cancel{(q+p)}~)^2$
should be neglected since no $\epsilon$ term can emerge. Thus
%
\begin{eqnarray}&& 
Tr ~\Bigl[\gamma_\chi ~ \Big \{
\cancel{(q-k)},~\cancel{(q-k)}~\gamma_\rho ~\cancel{q}~ \gamma_\sigma~
\cancel{(q+p)} \Big\}
\Bigr]
\nonumber\\&&
= 2 (q-k)^2 Tr ~\Bigl[\gamma_\chi ~\gamma_\rho ~\cancel{q}~ \gamma_\sigma~
\cancel{p} 
\Bigr]
\nonumber\\&&
 +2 (q-k)_\rho
Tr ~\Bigl[\gamma_\chi ~ 
\not\!k ~\cancel{q}~ \gamma_\sigma~
\not\!p  
\Bigr]
\nonumber\\&&
- 2 (q-k)q
Tr ~\Bigl[\gamma_\chi ~ 
\cancel{(p+k)}~\gamma_\rho ~ \gamma_\sigma~
\cancel{(q+p)} 
\Bigr]
\nonumber\\&&
 +2 (q-k)_\sigma
Tr ~\Bigl[\gamma_\chi ~ 
\not\!k ~\gamma_\rho ~\not\!q~
\not\!p  
\Bigr]
\nonumber\\&&
- 2 (q-k)(q+p)
Tr ~\Bigl[\gamma_\chi ~ 
\not\!k ~\gamma_\rho ~ \cancel{q} \gamma_\sigma
\Bigr] 
\label{LFE.11}
\end{eqnarray}
Now we  shift $q$
\begin{eqnarray}
q \longrightarrow q+r \qquad
r\equiv yk-xp + yp
\label{LFE.11.1}
\end{eqnarray}
and we drop all terms that are zero as a result of
the symmetric integration
%
\begin{eqnarray}&& 
Tr ~\Bigl[\gamma_\chi ~ \Big \{
\cancel{(q-k)},~\cancel{(q-k)}~\gamma_\rho ~\cancel{q}~ \gamma_\sigma~
\cancel{(q+p)} \Big\}
\Bigr]
\nonumber\\&&
=q^2 \frac{2}{D}\Bigg( 2 Tr ~\Bigl[\gamma_\chi ~\gamma_\rho ~\cancel{~(r-k)~}~ \gamma_\sigma~
\cancel{p} 
\Bigr]
 {
+ D  Tr ~\Bigl[\gamma_\chi ~\gamma_\rho ~\cancel{r}~ \gamma_\sigma~
\cancel{p} 
\Bigr]
}
\nonumber\\&&
  +
Tr ~\Bigl[\gamma_\chi ~ 
\cancel{k} ~\gamma_\rho~ \gamma_\sigma~
\cancel{p}
\Bigr]
- 
Tr ~\Bigl[\gamma_\chi ~ 
\cancel{(p+k)}~\gamma_\rho ~ \gamma_\sigma~
\cancel{(2r-k)}  
\Bigr]
\nonumber\\&&
 {-D(1-x)Tr ~\Bigl[\gamma_\chi ~ 
\cancel{k}~\gamma_\rho ~ \gamma_\sigma~
\cancel{p}  
\Bigr]}
 +
Tr ~\Bigl[\gamma_\chi ~ 
\cancel{k}~\gamma_\rho ~ \gamma_\sigma~
\cancel{p}  
\Bigr]
\nonumber\\&&
-
Tr ~\Bigl[\gamma_\chi ~ 
\cancel{k}~\gamma_\rho ~ \cancel{(2r+p-k)}\gamma_\sigma
\Bigr] 
 {
- D
Tr ~\Bigl[\gamma_\chi ~ 
\cancel{k}~\gamma_\rho ~ \cancel{r} \gamma_\sigma
\Bigr] \Bigg)
}
\nonumber\\&&
= q^2\frac{2}{D}\Bigg(2(y-1)  Tr ~\Bigl[\gamma_\chi ~\gamma_\rho ~\cancel{k}~ \gamma_\sigma~
\cancel{p} 
\Bigr]
 {
+ D y Tr ~\Bigl[\gamma_\chi ~\gamma_\rho ~\cancel{k}~ \gamma_\sigma~
\cancel{p} 
\Bigr]
}
\nonumber\\&&
  +
Tr ~\Bigl[\gamma_\chi ~ 
\cancel{k} ~\gamma_\rho~ \gamma_\sigma~
\cancel{p}  
\Bigr]
- (2x-1)
Tr ~\Bigl[\gamma_\chi ~ 
\cancel{p}~\gamma_\rho ~ \gamma_\sigma~
\cancel{k}  
\Bigr]
\nonumber\\&&
 {-D(1-x)Tr ~\Bigl[\gamma_\chi ~ 
\cancel{k}~\gamma_\rho ~ \gamma_\sigma~
\cancel{p}  
\Bigr]}
 +
Tr ~\Bigl[\gamma_\chi ~ 
\cancel{k}~\gamma_\rho ~ \gamma_\sigma~
\cancel{p}  
\Bigr]
\nonumber\\&&
- [ {2}(y-x)+1] 
Tr ~\Bigl[\gamma_\chi ~ 
\cancel{k}~\gamma_\rho ~ \cancel{p} \gamma_\sigma
\Bigr] 
 {
+ D(x-y)
Tr ~\Bigl[\gamma_\chi ~ 
\cancel{k}~\gamma_\rho ~ \cancel{p} \gamma_\sigma
\Bigr] \Bigg)
}
\nonumber\\&&
\simeq q^2\frac{2}{D} Tr ~\Bigl[\gamma_\chi ~\gamma_\rho ~\cancel{k}~ \gamma_\sigma~
\cancel{p} \Bigr]
\Bigg\{2(y-1) +yD-1 - (2x-1) 
\nonumber\\&&
+(1-x)D -1- 2(y-x)-1 +(x-y)D\Big\}
\nonumber\\&&
= q^2\frac{2}{D} Tr ~\Bigl[\gamma_\chi ~\gamma_\rho ~\cancel{k}~ \gamma_\sigma~
\cancel{p} \Bigr]\Big\{-4+ D \Big\}.
\label{LFE.12}
\end{eqnarray}
The factor $D-4$ is expected, since in four dimensions $\gamma_\chi$ 
anti-commutes with all $\gamma_\mu$ and therefore $T_{\rho\sigma}(k,p)$
is zero from start in eq. (\ref{LFE.8}).
\par
The integration over $x,y$ gives
\begin{eqnarray}&&
= q^2\frac{2}{D} Tr ~\Bigl[\gamma_\chi ~\gamma_\rho ~\cancel{k}~ \gamma_\sigma~
\cancel{p} \Bigr] {\frac{D-4}{2}}
\nonumber\\&&
=q^2\frac{ {D-4}}{D} Tr ~\Bigl[\gamma_\chi ~\gamma_\rho ~\cancel{k}~ \gamma_\sigma~
\cancel{p} \Bigr]
\label{LFE.12.1}
\end{eqnarray}
Now we multiply by 2 (the Feynman parameter), 2 (the crossed graph),
$-\frac{i}{(4\pi)^2}\frac{2}{D-4}$ from $q$ integration. We get
\begin{eqnarray}&&
2 \frac{i}{(4\pi)^2}~Tr ~\Bigl[\gamma_\chi ~\gamma_\rho
\not\!p ~ \gamma_\sigma~ \not\!k \Bigr].
\label{LFE.12.2}
\end{eqnarray}
Thus the results in eqs. (\ref{post.16}) and (\ref{LFE.12.2}) do
satisfy the LFE identity in eq. (\ref{LFE.7}).
%
\section*{Acknowledgements}
%
I  gratefully acknowledge the warm hospitality of the 
Department of Physics of the University of Pisa
and of the INFN, Sezione di Pisa.
\normalsize

\bibliography{reference}

\end{document}